\documentclass[preprint,12pt]{elsarticle}

\usepackage{amssymb}
\usepackage{multirow}
\usepackage{amssymb}
\usepackage{graphicx}
\usepackage{amssymb}
\usepackage{graphicx}
\usepackage{amsmath}
\usepackage{latexsym}
\usepackage[linesnumbered,ruled,algosection,nofillcomment]{algorithm2e}
\usepackage{subfigure}
\usepackage{epsfig}
\usepackage{floatflt}
\usepackage{rotating}
\usepackage{caption}
\usepackage{float}
\usepackage{amsfonts}
\usepackage{xcolor}
\usepackage{graphicx}
\usepackage{subfigure}
\usepackage{url}
\usepackage{longtable}
\usepackage{listings}
\usepackage{algorithmic}
\usepackage{amssymb}
\usepackage{setspace}
\usepackage{balance}
\usepackage{times,amsmath,epsfig,booktabs,multirow}
\usepackage{subfigure}
\usepackage{xcolor,paralist,times}
\usepackage{natbib}

\usepackage{epstopdf}

\journal{Journal of Systems and Software}

\begin{document}

\begin{frontmatter}
\title{A Novel Load Balancing Scheme for Mobile Edge Computing\tnoteref{label1} }
\tnotetext[label1]{This research is supported by the National Key Research and Development Program of China (2018AAA0103202); National Natural Science Foundation of China (62172322, 61751207, 61732013); Key Science and Technology Innovation Team of Shaanxi Province(2019TD-001).}
\cortext[cor1]{Corresponding author.}
\author[label2]{Zhenhua Duan}
\ead{zhhduan@mail.xidian.edu.cn}
\author[label2]{Cong Tian\corref{cor1}}
\ead{ctian@mail.xidian.edu.cn}

\author[label2]{Nan Zhang\corref{cor1}}
\ead{nanzhang@xidian.edu.cn}

\author[label3]{Mengchu Zhou}
\ead{zhou@njit.edu}
\author[label2]{Bin Yu\corref{cor1}}
\ead{yubin9011@126.com}

\author[label2]{Xiaobing Wang}
\ead{xbwang@mail.xidian.edu.cn}
\author[label2]{Jiangen Guo}
\ead{jgguo@mail.xidian.edu.cn}
\author[label2]{Ying Wu}
\ead{YingWu9208@126.com}

\fntext[label2]{Institute of Computing Theory and Technology, and ISN Lab, Xidian University, Xi'an, 710071, China}
\fntext[label3]{Department of Electrical and Computer Engineering, New Jersey Institute of Technology, Newark, NJ 07102, USA}

\begin{abstract}
 To overcome long propagation delays for data exchange between the remote cloud data center and end devices in Mobile Cloud Computing (MCC), Mobile Edge Computing (MEC) is emerging to push mobile computing, network control and storage to the network edges.
A cloudlet in MEC is a mobility-enhanced small-scale cloud, which contains several MEC servers located in close proximity to mobile devices.
The main purpose of a cloudlet is to stably provide services to mobile devices with low latency.
When a cloudlet offers hundreds kinds of services to millions of mobile devices,
it is critical to balance the loads so as to improve performance.

In this paper, we propose a three-layer mobile hybrid hierarchical P2P (MHP2P) model as a cloudlet. MHP2P performs satisfactory service lookup efficiency and system scalability as well as high stability. More importantly, a load balance scheme is provided so that the loads of MHP2P can be well balanced with the increasing of MEC servers and query load.
A large number of simulation experiments indicate that the proposed scheme is efficient for enhancing the load balancing performance in MHP2P based MEC systems.
\end{abstract}

%


\end{frontmatter}


\section{Introduction}
Recently, mobile devices (such as smart phones, tablet computers etc.) are becoming more useful tools for learning, entertainment, social networking, updating news and businesses.
In the beginning, mobile users do not get the same satisfaction compared with desktop due to resource limitations of mobile devices.
With the tremendous advancements in wireless communications and networking, Mobile Cloud Computing (MCC)  \cite{fernando2013mobile,noor2018mobile,bharati2021review}
is emerging as a new paradigm of computing in the last decade to improve above scenario. With MCC, the high-rate and highly-reliable air interface allows resource-constrained end-user devices to offload some computations to the remote resourceful cloud. After the evolution of MCC, many cloud computing services such as mobile health care, mobile learning and mobile gaming can be directly accessible from mobile devices \cite{hameed2019application,rimale2016survey,cai2013next,Ref-Qiao}.

With the advent of cloud computing, the back-end server is typically hosted at the cloud data center. Though the use of a cloud data center offers various benefits such as scalability and elasticity \cite{Ref-Li}, its consolidation and centralization lead to a large separation between a mobile device and its associated data center.
Offloading computation to the public cloud may involve long latency and low bandwidth for data exchange between the public clouds and edge device through the Internet \cite{satyanarayanan2009case,cho2016survey}.
Thus, MCC is not adequate for a wide-range of emerging mobile applications that are latency-critical.

As a new platform proposed by the \emph{European Telecommunications Standard Institute} (ETSI) in 2014, Mobile Edge Computing (MEC) ``provides IT and cloud-computing capabilities within the Radio Access Network (RAN) in close proximity to mobile subscribers" \cite{ETSI}.
In an MEC platform, the edge responsibility is increased.
Computation and service are allowed in cloudlets at the edge, which include several proximal MEC servers.
Mobile applications benefit from MEC by offloading their computation-intensive tasks to the MEC servers, aiming to reduce the network latency and bandwidth consumption \cite{mao2017survey,ren2019survey,zhang2018mobile}.

There are three basic components in the model of MEC: (1) mobile devices include all types of devices (both mobile phones and IoT devices) connected to the network; (2) MEC servers are typically small-scale data centers deployed in close proximity with end users. MEC servers have the responsibility of traditional
network traffic control (both forwarding and filtering) and hosting various mobile edge applications (edge health care, smart
tracking etc.) and (3) public cloud is the cloud infrastructure hosted in the Internet.

Since the lower requirements for computation power and storage space, more devices can act as MEC servers. Harvesting the vast amount of the idle computation power and storage space distributed at the network edges can yield sufficient capacities for performing latency-critical tasks at mobile devices.
Models of computation tasks, communications, mobile devices and MEC servers are studied in recent years. These state-of-the-art researches involve deployment of MEC systems \cite{cui2021survey}, mobility management for MEC \cite{mehrabi2019survey} and security-and-privacy issues in MEC \cite{roman2018mobile} etc.
However, there are still challenges in a cloudlet including millions of MEC servers and mobile users:

\begin{itemize}
\item[(1)] In a geographic area, there may exist a large number of MEC servers which can supply kinds of services for enormous amount of mobile users. For a mobile user, how to find out a suitable MEC server efficiently and then connect it is a concerned issue \cite{lee2019low};
\item[(2)] With the popularity of MEC, more and more network elements at the network edge are willing to act as an MEC server. To ensure stability of a cloudlet, the leaving and joining of MEC servers should be detected in time. For key servers which manage others, recovery approaches should be performed to continue the MEC service \cite{satria2017recovery};
\item[(3)] Unlike remote cloud data centers, the computation power and storage space in an MEC server is insufficient to serve too many mobile users simultaneously. For a cloudlet with several MEC servers, it is necessary to ensure performance of the service by applying load balancing schemes to managing MEC servers in the cloudlet \cite{zhang2020secure}.
\end{itemize}

In order to construct a cloudlet performing satisfactory server lookup efficiency and system
scalability as well as high stability, we integrate a three-layer mobile hybrid hierarchical P2P (MHP2P) network as a cloudlet into MEC systems. It can take advantages of both Distributed Hash Table (DHT) and flooding methods to improve performance \cite{duan2016two}. MHP2P uses Chord \cite{Chord2001} as the upper layer, clusters as the middle one and mobile devices as the lower one.
Here the whole system is regarded as a Chord ring composed of a set of virtual nodes. Each virtual node corresponds to a cluster which is a group of MEC servers in the close geographical area.

In our system, an MEC server can be any device which can access network, including workstations, desktop computers and tablet
computers. In practice, it can be a hospital, a market or a petrol station offering online service, or idle computers willing to provide computing service.
MHP2P is good at efficiency because a search message is transferred within DHT on the top level in a large scale and
flooding search is limited to clusters each of which consists of a small number of MEC servers only. MHP2P is stable because MEC servers joining or leaving the system is limited to a cluster. MHP2P has better scalability due to the combination of DHT and flooding approaches. It inherits the good scalability from DHT since clusters are organized into a Chord ring. Meanwhile, inter-cluster and intra-cluster load balancing schemes are provided to solve the problem of load imbalance in the MHP2P upper and middle layers.

The contributions of the paper are summarized as follows:
\begin{itemize}
\item[(1)] A novel cloudlet model called MHP2P is presented for MEC systems. It has high stability, scalability and look-up efficiency. More importantly, load balancing schemes are provided so that the loads in MHP2P can be well balanced with the increasing of MEC servers and query load;
\item[(2)] Except for theoretical analysis, a large number of experimental simulations are conducted and the results show that the proposed schemes can significantly improve the degree of load balancing even the model size is in millions.
\end{itemize}

The paper is structured as follows. A motivating example is given in the next section. Section \ref{sec:cloudletmodel} provides our novel cloudlet model.
Section \ref{sec:Loadbalancing} describes the details of the inter-cluster and intra-cluster load balancing schemes.
A large number of simulation results are shown in Section \ref{sec:Simulation}. Finally,  conclusion is drawn in Section \ref{sec:conc}.

\section{A motivating example}\label{sec:motivating}


In our daily life, health  communities and city hospitals provide emergency services. An  emergency department usually provides longer service time than other departments, even remains open 24 hours. With the development of Internet,
more and more emergency departments offer audio or video services for medical advice and diagnosis online.
Based on the IP address, an ambulance can find the precise geographical location at the time when a patient requires.

When a patient expects to obtain some emergency medical services, he will first open an application installed on his mobile device.
Then the application will send the query to the data center and receive information about suitable ones.
In the case in which all related information is stored in the remote cloud, the long propagation distance from the end user to the remote cloud center will result in excessively long latency.
What is worse, the increasing of query number will lead to the network congestion, in which situation the response takes more time.
Suppose a patient needs service supplied by a emergency department, long time latency may make the patient miss the best
time for rescue.

From the above scenario, it can clearly be seen that MCC is unsuitable for latency-critical service. Instead, an MEC can be employed to avoid frequently requesting the remote cloud for each query. Further, a cloudlet consisting of several MEC servers in proximity to mobile users is really necessary. Related information about service suppliers is stored in MEC servers.
In fact, service suppliers can act as MEC servers since they are constantly online and have enough storage capability. With an ideal environment, sufficient service suppliers can join together in a cloudlet.

With the increasing of MEC servers and service queries, there are three problems that we must take into account:

\begin{itemize}
\item[(1)] It is necessary to store information about different kinds of service suppliers in a specific rule which allows one to efficiently look up the required information;
\item[(2)] MEC servers should be well managed so that the failure of an MEC server should be detected in time. A cloudlet should maintain stability even when MEC servers frequently join and leave;
\item[(3)] To avoid network congestion and the overload of an MEC server, the loads of the whole cloudlet should be well balanced to ensure its performance and fairness.
\end{itemize}

\section{A novel cloudlet model}\label{sec:cloudletmodel}

This section presents the architecture of MHP2P as a novel cloudlet model. MHP2P is a hybrid hierarchical P2P network combining both unstructured and structured P2P networks. The upper structured layer, which stands for the overall framework, is based on Chord using DHT. The middle unstructured layer, including several MEC servers, is based on a cluster using flooding. The lower layer are mobile devices which request services.

\subsection{MHP2P framework}

\begin{figure}[!htb]
  \centering
  \includegraphics[width=120mm]{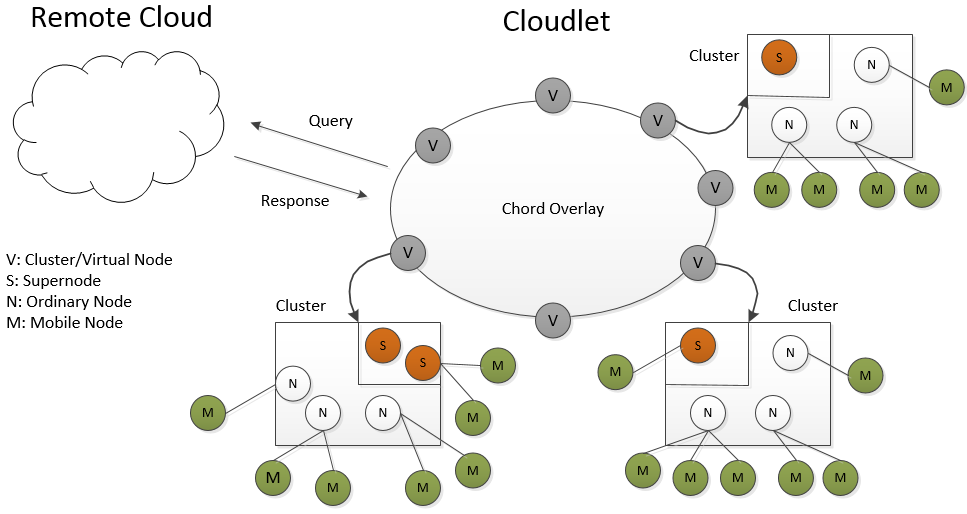}
  \caption{Framework of MHP2P}
  \label{fig:MHP2P}
\end{figure}

The framework of MHP2P is shown in Fig. \ref{fig:MHP2P}. It evolves from HP2P architecture
presented in \cite{duan2016two}.  As you can see, a node (N) stands for an MEC server in a cloudlet, which can be workstations, desktop computers, tablet computers etc.
An MEC server has two functions: (1) to supply its own service and (2) to store metadata for other service suppliers. A piece of metadata is used to record the necessary information of a service supplier, including the service name, geography location, IP address, business hours, and service description etc. Nodes are classified into two types: Ordinary node (ODN) and Supernodes (S). Supernodes are the MEC servers with high network bandwidth, computation power and shared disk storage capacity. In general, a supernode will perform some specific  tasks beyond ordinary ones. A cluster (CL) is a group of MEC servers in a geographical area. The same style of management and routing is followed in a cluster. A virtual node (V) is actually a cluster. However, in the hierarchical networks, a cluster appears at the top level as a virtual node. A mobile device (M) is a service requester, which is willing to connect a service supplier in time to obtain required service.

When a mobile device needs some service, it initiates a request to a boot server. According to the response from the boot server, the mobile device then connects with a suitable MEC server to publish its service query. After the look-up process in MHP2P, the mobile device finally gets a successful response if the service can be found in the cloudlet. Otherwise, the query will be sent to the remote cloud, which is out of scope of this paper.

MEC servers in our model are organized in clusters where messages are forwarded by Gossip Flooding \cite{voulgaris2003robust}. In each cluster, some supernodes are elected from ordinary nodes. All clusters are organized in a Chord mechanism. Each cluster is as a virtual node in the Chord ring that has a routing table maintained by its supernodes. Communications among clusters are forwarded by their supernodes.

A candidate will substitute for a supernode if the supernode leaves its cluster.  Only when no node exists or all supernodes collapse at the same time, will a cluster disappear. Hence, a cluster tends to be stable in a long period since its establishment.  The Chord overlay routing information will not be influenced by the frequently joining or leaving of nodes, which will only impact on several neighborhood nodes in a cluster. Therefore, the whole network keeps stable. Further, the nodes with high bandwidth are effectively utilized with the employ of ``power'' supernodes. Finally, by combining the high search efficiency and scalability of DHT, the whole system performance can be significantly enhanced.

\subsection{Architecture of upper layer}

The original protocol of Chord supports just one operation: mapping a given key onto a node. In MHP2P, the original Chord is modified to map one dimensional hash space onto a cluster.
The key is a service name and the value is IP address where the metadata is stored.
Clusters in the modified Chord are identified by a consistent hash function \cite{karger1997consistent}. When a cluster is established, the IP address of its supernode is hashed to be an m-bit identifier by a hash function. With the same hash function, a service name is hashed to be m-bit ID. Then the relation between service and clusters is established \cite{duan2016two}.

A successor item of the finger table in the upper layer of MHP2P points a supernode table recording a group of supernodes in the same cluster.
Given a service name as a key, the Chord overlay is able to find the cluster that is responsible for storing the key's value by a Chord routing mechanism. In a cluster, all supernodes maintain a finger table together. The supernode table of related clusters  is updated when a supernode is changed. In a supernode table, it must be assured that the cluster can be valid if at least one supernode item is valid. Therefore, the existence of several supernodes in a supernode table enhances the stability of the routing table.

\subsection{Architecture of middle layer}

A cluster in MHP2P is used as the middle layer to restrict the flooding extension and enhance the system stability.
MEC servers with vicinal geography locations are organized into a cluster.
There are two kinds of nodes in a cluster: supernodes and ordinary nodes. Nodes with high capability act as the leaders of clusters, which are called supernodes. Supernodes collaborate to manage the cluster and maintain its finger table. A cluster is established when the first node joins it, and the first node's ID is considered to be the cluster's ID.
At the same time, the first node becomes the first supernode in it. The nodes joining it afterward are generally considered to be ordinary nodes. Other supernodes can be selected by the cluster establisher.
All supernodes work together to manage the cluster to transfer the information across different clusters.

In one cluster, the main way for the communication of different nodes is a lightweight flooding in which a node just sends a newly generated message to a set of randomly selected neighbor nodes.
These nodes do similarly in the next round, and so do other nodes until the message is spread to the entire cluster
or TTL is reduced to zero.

\subsection{Architecture of lower layer}

The lower layer of our model is composed of mobile devices and the number of devices can reach millions. The application installed on a mobile device sends a message to a boot server first.
After that, the boot server will establish a connection between the mobile device and its nearest MEC server.
Then a query is sent from the mobile device to the MEC server, which contains the expected service name and IP address of the mobile device. The MEC server will help the mobile device forward the service query to the Chord.
The detailed process to look up the service metadata will be given in the next subsection.

After one or more service suppliers are located, the MEC server connecting with the mobile device will first get the query result. The nearest service supplier will be selected by the MEC server and then sent to the mobile user. At last, the mobile user connects with the service supplier directly in peer-to-peer fashion.

\subsection{Lookup service of MHP2P}
After a query is transmitted from a mobile device to its nearest MEC server, it is sent to the Chord to locate the cluster that stores the metadata, and then queries are spread in the cluster through a lightweight flooding search.

The detailed search process of MHP2P is as follows:

\begin{itemize}
\item[1.] A mobile device M sends a query to its nearest node P$_i$;
\item[2.] P$_i$ in CL$_i$ sends the query to S$_i$ in CL$_i$;
\item[3.] S$_i$ finds a successor (a cluster CL$_j$) through the Chord;
\item[4.] S$_i$ forwards the query to successor CL$_j$;
\item[5.] In CL$_j$, its supernode S$_j$ floods the query;
\item[6.] When found, node P$_j$ in CL$_j$ returns a successful response to P$_i$, go to 7; and otherwise P$_i$ sends the query the remote cloud;
\item[7.] After the filtering procedure, P$_i$ sends suitable information to M.
\end{itemize}

\section{Load balancing in our model}\label{sec:Loadbalancing}

%
%



The MHP2P upper layer is constructed by Chord which employs protocols based on DHT, thereby leading to a serious load imbalance problem that some clusters initiate flooding queries in high frequency while others are in idle states. In the MHP2P middle layer which uses the Gossip protocols, the load imbalance causes the problem that some nodes maintain too many metadata that consume too much of their processing capacity, while others maintain few metadata only.

To solve the load imbalance problem in MHP2P, this section presents a load balancing scheme, which consists of inter-cluster and intra-cluster load balancing. In the former, we achieve the MHP2P upper layer load balance by means of  cluster splitting and cluster moving to adjust loads of nodes on the Chord ring. In the latter,  the MHP2P middle layer reaches load balance by making full use of supernodes. Since the supernodes own the whole view of their corresponding cluster, their guidance can be used to enable a node with heavy load to find another with light load in the same cluster easily. For clarity, Table \ref{tt1} lists the terms and notions used in this section.

\begin{table}
\begin{center}
\caption{Terminology}\label{tt1}
{\small\begin{tabular}{|c|l|c|c|c|} \hline
\textbf{Notation} & \textbf{Definition} \\
\hline
$Cluster_{A}$ & cluster $A$ \\
\hline
$Length_{A}$ & length of address space that $Cluster_{A}$ should manage on the Chord ring \\
\hline
$Load_{A}$  & the load of cluster $A$\\
\hline
$load_{n}$  & the load of node $n$\\
\hline
$capacity_{n}$  & the power of node $n$\\
\hline
$Load_{avr}$  & the average cluster load by estimation\\
\hline
$rate_{a}$  & the load rate of node $a$\\
 \hline
$rate_{avr}$  & the average load rate of a cluster\\
 \hline
$\alpha$  & $\alpha \in (1,2)$ is used to determine whether a node is with high load rate or low load rate \\
\hline
 \multirow{2}{*}{$\beta$} & $\beta \in [0,0.5)$ is used to determine whether a cluster can find a neighbor cluster to \\
 &transfer or receive some loads\\
\hline
$\gamma$ & $\gamma \geq 2$ is used to determine whether a cluster can split to reduce its cluster loads\\
\hline
\multirow{2}{*}{$k$} & $k$ is used to determine the number of clusters from which a cluster should  receive \\
&the load messages to estimate $Load_{avr}$ \\
\hline

\end{tabular}}
\end{center}
\end{table}

\subsection{Inter-cluster load balancing}

\subsubsection{Load definition in inter-cluster load balancing}

 The load of a node in a network can represent the utilization of bandwidth, the utilization of CPU cycles, the number of items to maintain and the number of messages to process. In the MHP2P architecture, each node completes most of operations by sending and receiving messages such as flooding messages and metadata maintaining messages. Thus we define  load as the number of messages a node to process in each period of time. Cluster load is defined in terms of  the average number of messages  in the cluster for a node to process:
$$Load_{A}=\frac{1}{N
um_{A}} \sum_{n\in A}^{}load_{n}~~~~~(1)$$
where $Num_{A}$ is the number of nodes in cluster $A$, and $n$ is a node index.

\subsubsection{Average cluster load estimation}

Since the MHP2P upper layer is a fully distributed network, we are not able to obtain its average cluster load directly. In its  inter-cluster load balancing, a supernode in each cluster selects $k \times log_{}N$ other clusters randomly to obtain their cluster loads at regular intervals. Finally, a supernode estimates the average cluster load by calculating the average cluster load of $k \times log_{}N$ clusters as follows:
$$Load_{avr}=\frac{1}{k \cdot log_{}N} \sum_{m=0}^{k \cdot log_{}N}Load_{m}~~~~~(2)$$
where $N$ represents the number of nodes available in the network.
Because the inter-cluster load balancing scheme does not need high precision in an average cluster load estimation, we should not set $k$ to a high value. According to our
 simulations, if the network is in a low degree of the inter-cluster load balancing, $k$ is set to 4. If it is in a high degree of the inter-cluster load balancing, $k$ is set to 1. In this way, the estimation of the average cluster load is in the interval [$0.85\cdot Load$, $1.15\cdot Load$] with high probability, where $Load$ is the average cluster load of the entire network in practice.

\subsubsection{Principle of inter-cluster load balancing}

Since the number of messages in each cluster cannot be controlled directly, we achieve the adjustment of the number of messages by controlling the number of metadata in each cluster indirectly. In MHP2P, a supernode caching mechanism can avoid hot issues, such that  the number of metadata in each cluster is proportional to the number of messages on the whole. In other words, the more metadata a cluster has the more messages it processes. This is because if a cluster has more metadata, it should have a higher probability to be retrieved, and also cause more messages to maintain metadata.

In the upper layer Chord ring, each supernode of clusters is required to send its load balance request to its
 neighbors from time to time. Let $Load_{avr}$  denote the average load of all clusters in the Chord ring
 and $L_r$ denote the requesting cluster's load at present. If  $L_r < Load_{avr}$, the requesting cluster is a light load cluster;
  if  $L_r = Load_{avr}$, it is a moderate cluster;
  if $L_r > Load_{avr}$,  it is a heavy load cluster; and further if $L_r \geq \gamma \cdot Load_{avr}$ with a threshold value $\gamma$, it is a very heavy cluster. To balance loads among clusters, the basic idea is: a) to move some loads of a heavy load cluster to a light load one; and b) split a cluster with very heavy load into two whenever a supernode is requesting.
\begin{spacing}{1.5}
(a) Moving strategy
\end{spacing}
 To move loads reasonably between clusters, we set a threshold value $V_m$ such that a light cluster $r$
 can acquire some loads from a heavy one $x$ if $L_r < V_m \cdot L_x$ or a heavy one $x$ can
 release some loads to a light neighbor cluster $r$ if $V_m \cdot L_x \geq \cdot L_r$.
 Roughly speaking, the distance from a cluster to its predecessor cluster in a counterclockwise direction can be
 treated as its load. Therefore, if a light load cluster intends to acquire extra load from a heavy successor cluster,
 it needs to move in a clockwise direction closer to its successor cluster while if a heavy load cluster plans to release some
 loads to its light predecessor cluster, it needs to move in a counterclockwise direction  closer to its predecessor
 cluster so as to change the distribution of metadata between a cluster and its successor cluster.

In order to balance loads precisely, on one hand, we need to know the percentage of loads of a cluster to be released; on the other hand, we need to estimate the exact  distance a cluster will move clockwise or counterclockwise.
For instance, if a cluster is requested to have equal load with its successor cluster, the load of the heavier cluster can decrease $\beta$ ($\beta \in [0,0.5)$) times of its load by moving clockwise or counterclockwise through a small amplitude. This is an effective load transfer between neighbor clusters. If $\beta$ is $0.2$, a load transfer can make a heavier cluster decrease 20 percentage of its load. In MHP2P, if two adjacent clusters are $Cluster_{A}$ and $Cluster_{B}$, and their loads  $Load_{A}$ and $Load_{B}$ satisfy:
$$Load_{B}\geq\frac{Load_{A}}{1-2\cdot\beta}~~~~~(3)$$
 $Cluster_{A}$ can obtain some metadata to decrease $\beta$ times of $Load_{B}$.
In  this case,  we set $V_m = 1-2\cdot\beta$. Since the supernode does not know every ID of the metadata in its cluster, we cannot determine the exact length a cluster should move on the Chord ring with low cost. Thus, we assume that IDs of metadata are evenly distributed in each  address space of the cluster. In this way, we can estimate the $Length$ a cluster should move by
$$Length=\frac{(Load_{B}-Load_{A})\times Length_{B}}{2\times Load_{B}}~~~~~(4)$$

To fully clarify the load transfer between two adjacent  clusters, we have two cases, where $Cluster_{B}$ is assumed to be the successor of $Cluster_{A}$:

Case1: $Load_{B}\geq\frac{Load_{A}}{1-2\cdot\beta}$, $Cluster_{A}$ moves clockwise through $$\frac{(Load_{B}-Load_{A})\times Length_{B}}{2\times Load_{B}}$$ on the Chord ring, while  $Cluster_{B}$  transfers the corresponding Chord ring regions of metadata to $Cluster_{A}$.

Case2: $Load_{A}\geq\frac{Load_{B}}{1-2\cdot\beta}$, $Cluster_{A}$ moves counterclockwise through $$\frac{(Load_{A}-Load_{B})\times Length_{A}}{2\times Load_{A}}$$ on the Chord ring, while $Cluster_{B}$ transfers the corresponding Chord ring regions of metadata to $Cluster_{A}$.

\begin{spacing}{1.5}
(b) Splitting strategy
\end{spacing}

If load $L_r$ of a requesting cluster satisfies $L_r \geq \gamma
\cdot Load_{avr}$ with a threshold value $\gamma$ ($\gamma \geq 2$), its load
is too heavy. In this case, a splitting
strategy is employed \cite{duan2016two}. To do so,  a new cluster node called
$y$ is created in the counterclockwise direction with a suitable
distance from the cluster $r$, such that some of $L_r$ can be moved to the
new cluster $y$. As for how to split a cluster, it is out of scope of this
paper. Please refer to \cite{duan2016two} for details.

The basic idea of the inter-cluster load balancing is to split clusters and balance the load among neighbor  clusters (the successor  and predecessor  of a cluster). In this way, we reduce the load imbalance among clusters according to the individual topology of the MHP2P upper layer network.

\subsubsection{Analysis of inter-cluster load balancing}

In the inter-cluster load balancing mechanism of MHP2P, the upper layer network can achieve load balance within a narrow range of the Chord address space by cluster moving and splitting. New nodes continually joining the clusters with heavy load will lead to a high probability to split the latter. As the inter-cluster load balancing proceeds, the number of clusters  increases through the splitting in the heavier load area of the upper layer Chord address space. Thus  the load balancing in the whole MHP2P upper layer network is achieved.

As the supernode of each cluster needs to collect the loads of other $k\cdot log_{}N$ clusters to estimate the average load of all clusters, it needs to send $k\cdot log_{}N$ additional messages to obtain load information of these $k\cdot log_{}N$ clusters in every cycle. Note that only one supernode is  in charging a cluster at any time though there may be a few supernode candidates within a cluster.

\subsection{Intra-cluster load balancing}\label{sec4}
Intra-cluster load balancing is intended to solve the problem of
load imbalance in the MHP2P middle layer network by the supernode in
each cluster to schedule the loads in terms of computing power of its nodes.

\subsubsection{Load rate definition in intra-cluster load balancing}

According to our  simulations, the load of each node mainly comes from flooding query messages, metadata maintenance and network topology maintenance.
The load rate of a node can be defined as:
$$rate_{a}=\frac{load_{a}}{capacity_{a}}~~~~~(5)$$

However, for each node in the same cluster, various numbers of messages are mainly from  metadata maintenance. The number of metadata maintenance messages for a node is determined by the number of metadata items this node should maintain. In MHP2P, the mechanism of the topology maintenance can guarantee that each node in the cluster has roughly the same number of neighbors so as to keep the balanced distribution of nodes in the flooding network. Moreover, we use an improved  flooding strategy based on RBFS (Random breadth-first search) which can increase the randomness of the process in the flooding.
These two mechanisms can keep  the flooding query messages  and network topology maintenance messages  roughly the same  for each node in the same cluster. Hence, the load of each node can be adjusted by numbers of metadata items.

\subsubsection{Principle of intra-cluster load balancing}
   To balance the loads among nodes within a cluster, the supernode maintains a hash table H and a sorted link list L.
   Each record in L contains load rate, number of metadata  to be removed and time stamp of a heavy node $A$ ($rate_{A} > \alpha\cdot rate_{avr}$) in descending order while each record in H contains  node ID, and a pointer pointing to a relative record in L.  Whenever a node requires to release some load (($rate_{A}-rate_{avr}$)$\cdot capacity_{A}$),  the supernode makes a record in H and its relative record in L. When a light load node $B$ ($rate_{B} < $($2-\alpha$)$\cdot rate_{avr}$) requests to acquire some load (($rate_{avr}-rate_{B}$)$\cdot capacity_{B}$), the supernode checks list L and allocates sufficient  metadata from suitable heavy nodes to it, and then updates H and L.

\subsubsection{Analysis of intra-cluster load balancing}

In the intra-cluster load balancing scheme, we make full use of the feature that there are several supernode candidates in each cluster. Because a supernode has a better view of the whole cluster, with its help, a node with low load rate can find the node with high load rate efficiently.

In order to keep the load information for a node with high load rate, a supernode should  manage a hashtable and  sorted link list. The size of the list is determined by the number of the nodes with high load rate. However, it does not cost too heavy load for the supernode, because the total number of nodes in a cluster is small. As the network evolves, the load is more and more evenly distributed, and  the number of nodes with high load rate decreases quickly.

\section{Simulation results for load balancing}\label{sec:Simulation}

In this section, we analyze the very significant performance improvements that are
owing to the proposed algorithms by simulations. For fairness, a third-party simulation engine PeerSim \cite{peersim} is used. PeerSim provides an engine and two simulation models: cycle and event-driven models. Table II lists parameters of the simulation environment for our algorithms, and the values we set unless otherwise specified.

We run each trial of the simulation for 50 cycles. According to statics of Kademlia \cite{steiner2007global}, the maximum number of nodes is three times the minimum number of nodes. Thus we set the proportion of nodes arrival to departure 1\%:1\%, 1\%:3\%, 3\%:1\% such that the number of nodes is kept the same, decreasing to roughly one-third, or increasing to 3 times, in 50 cycles. Finally, in order to indicate the reliability of the simulation results, each selected data point in our plots represents the average simulation result over 5 trials. Under $2^{20}$ nodes simulation scale, each trial takes about 10 hours. Table \ref{tt2} and \ref{tt3} list the parameters of our simulations, and the values we set unless otherwise specified.

\begin{table}
\begin{center}
\caption{Simulation Environment}\label{tt2}
{\small\begin{tabular}{|c|c|c|c|c|} \hline
\textbf{Environment parameters} & \textbf{Default values}\\
\hline
Networksize & $2^{16}$\\
\hline
Ratio of node arrival:departure & 1\%:1\% 1\%:3\% 3\%:1\%\\
\hline
Number of metadata for each node & $10$\\
\hline
Node capacity distribution  & Pareto: shape $2$\\
\hline
Simulation cycles  & $50$\\
\hline
\end{tabular}}
\end{center}
\end{table}

\begin{table}
\begin{center}
\caption{Algorithm Parameters}\label{tt3}
{\small\begin{tabular}{|c|c|c|c|c|} \hline
\textbf{Algorithm parameters} & \textbf{Default values}\\
\hline
$\alpha$  & $1.4$ \\
\hline
$\beta$  & $25\%$\\
\hline
$\gamma$  & $2.0$\\
\hline
$k$  & 4 in first 10 cycles, 2 in 11 to 20 cycles, 1 in last 30 cycles \\
\hline
\end{tabular}}
\end{center}
\end{table}

\subsection{Simulation of the inter-cluster load balancing algorithm}

In simulations, we mainly calculate the ratio of the load of the cluster with the heaviest load and average cluster load in order to determine whether the inter-cluster load balancing algorithm can solve the problem that some clusters have too heavy load. In order to determine whether the algorithm has effect on improving the load balance of the whole MHP2P upper layer network, we calculate the relative standard deviation of the cluster load.

\subsubsection{Simulations under different network sizes}

\begin{figure*} \centering \subfigure[The ratio of the heaviest cluster load  and average cluster load under different network sizes.] { \includegraphics[width=5cm,height=4.2cm]{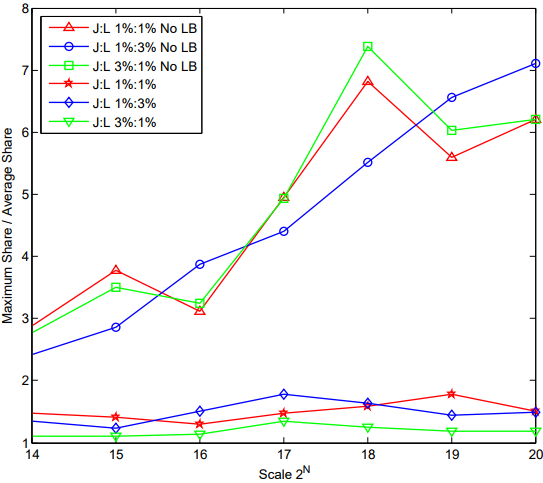} \label{cluster1maxload} } \ \ \subfigure[The standard deviation of cluster loads under different network sizes.] { \includegraphics[width=5cm,height=4.2cm]{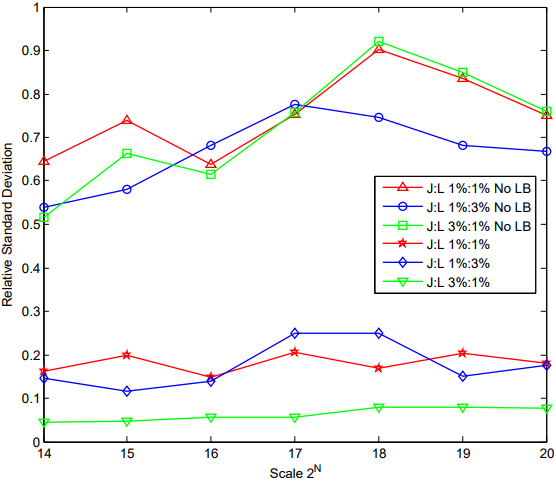} \label{cluster1standaredeviation} }
\subfigure[The ratio of the heaviest cluster  load  to average cluster load under different numbers of metadata items.] { \includegraphics[width=5cm,height=4.2cm]{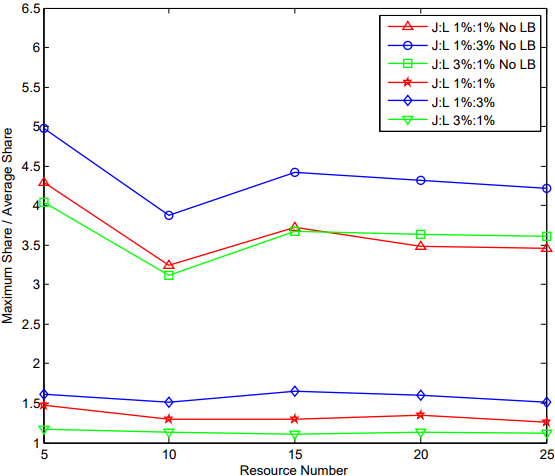} \label{cluster2maxload} }\ \
\subfigure[The standard deviation of the cluster load number of metadata items under different numbers of metadata items.] { \includegraphics[width=5cm,height=4.2cm]{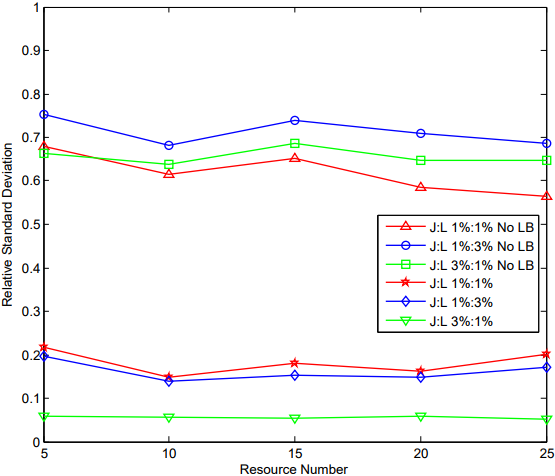} \label{cluster2standaredeviation} }

 \subfigure[The ratio of the heaviest cluster load  to average cluster load under different node request rates.] { \includegraphics[width=5cm,height=4.2cm]{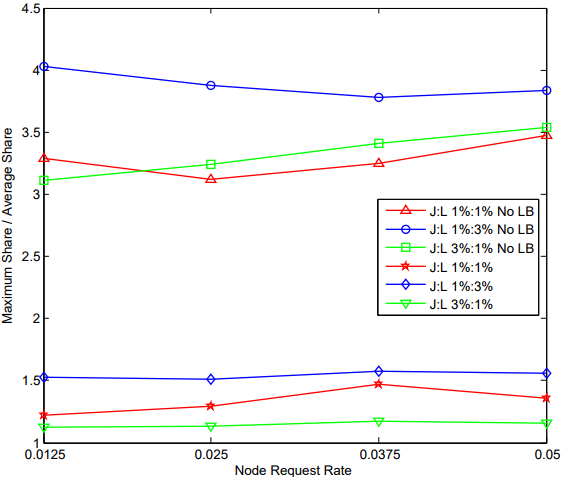} \label{cluster3maxload} }\ \ \subfigure[The standard deviation of the cluster load numbers of metadata items under different node request rates.] { \includegraphics[width=5cm,height=4.2cm]{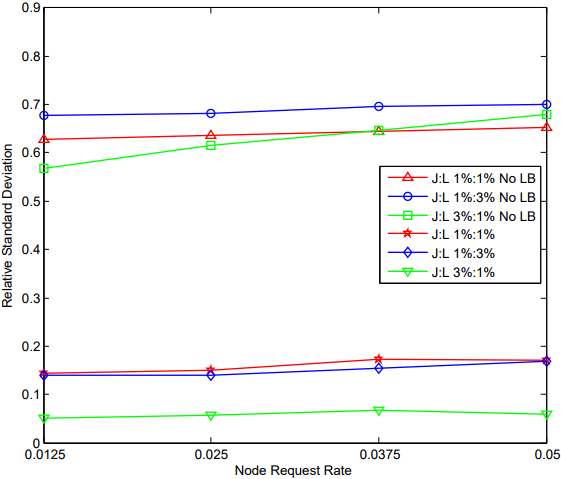} \label{cluster3standaredeviation} } \caption{Simulations for the inter-cluster load balancing} \label{cluster1standaredeviationandmaxload} \end{figure*}

Since the number of clusters and size of each cluster are both determined by the size of the whole network, we do simulations under different network sizes to determine whether the inter-cluster load balancing algorithm fits for MHP2P under different numbers of clusters and sizes of clusters. The network size is from $2^{14}$ to $2^{20}$. Thus,  the largest network size is more than one million nodes.

Fig. \ref{cluster1maxload} shows that the inter-cluster load balancing algorithm can always keep the ratio of the load of the heaviest load cluster  to average cluster load below 2 under different network sizes. By contrast, the ratio without the inter-cluster load balancing algorithm in MHP2P ranges from 3 to 7. In Fig. \ref{cluster1standaredeviation}, the standard deviation of the cluster load with the inter-cluster load balancing algorithm in MHP2P is much smaller than that without it. In conclusion, the inter-cluster load balancing algorithm can make loads in a balanced distribution among clusters under different network sizes.

\subsubsection{Simulations under different number of metadata items}

As mentioned earlier, the metadata maintenance message is also an important part of the total messages of an MHP2P network. When the network has a different number of metadata items, the metadata maintenance messages are in different proportions of the total messages. Thus we do this simulation to determine whether the inter-cluster load balancing algorithm has effect on the MHP2P network with different proportions of the metadata maintenance messages. The average number of metadata for each node is 5, 10, 15, 20 and 25.

Figs. \ref{cluster2maxload} and \ref{cluster2standaredeviation} show that the inter-cluster load balancing algorithm can not only sharply decrease the load of the heaviest load cluster but also improve the degree of the load balance of the whole MHP2P network under different numbers of metadata items. In fact, with $2^{19}$ nodes scale, the average load and heaviest load of a cluster is respectively 423 and 2811 items without using the algorithm while 362 and 718 with  the algorithm.

\subsubsection{Simulations under different node request rates}

Since different node request rates  lead to different proportions of flooding messages in total messages of an MHP2P network, this simulation is used to indicate if the algorithm has effect on the MHP2P network under different numbers of flooding messages. In the simulation, we vary the node request rate between 0.0125 and 0.05.

Figs. \ref{cluster3maxload} and \ref{cluster3standaredeviation} show the ratio of the load of the heaviest load cluster to average cluster load keeps below 1.5 and the degree of the load balance of the whole MHP2P network is greatly improved with the inter-cluster load balancing algorithm.

\subsubsection{Simulations under different settings of parameters $\beta$ and $\gamma$}

\begin{figure*} \centering \subfigure[The relation among $\beta$, $\gamma$ and the ratio of the heaviest  cluster load  and average cluster load when the network size is not subject to quick changes.] { \includegraphics[width=5cm,height=3.8cm]{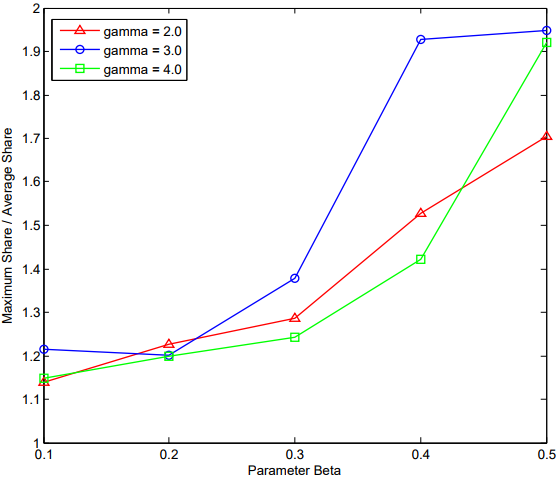} \label{cluster4maxload1} } \ \ \subfigure[The relation among $\beta$, $\gamma$ and the standard deviation of the cluster load when the network size is not subject to quick changes.] { \includegraphics[width=5cm,height=3.8cm]{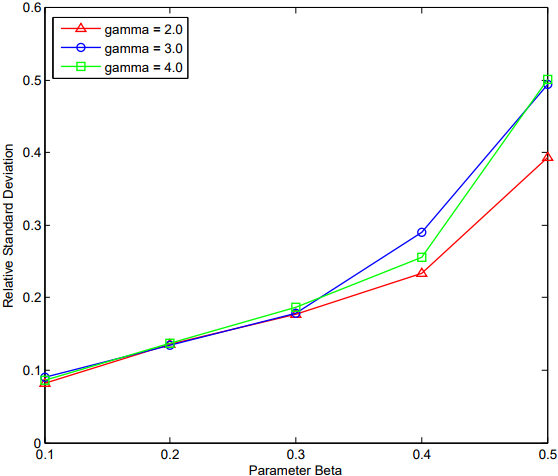} \label{cluster4standaredeviation1} }

\subfigure[The relation among $\beta$, $\gamma$ and the number of metadata movement when the network size is not subject to quick changes.] {\includegraphics[width=5cm,height=3.8cm]{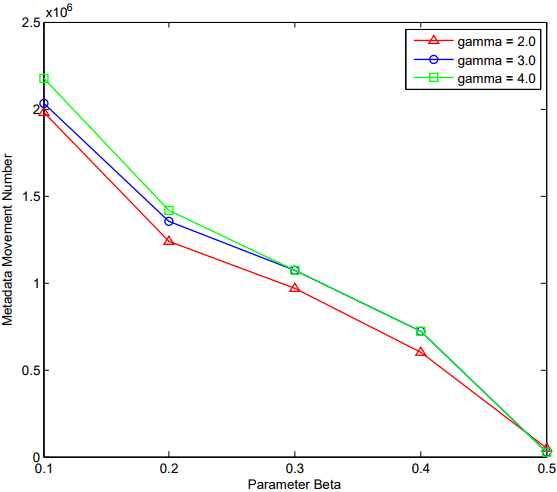} \label{cluster4move1} }\ \
\subfigure[The relation among $\beta$, $\gamma$ and the ratio of the heaviest cluster  load  and average cluster load when the network size is decreasing quickly.] { \includegraphics[width=5cm,height=3.8cm]{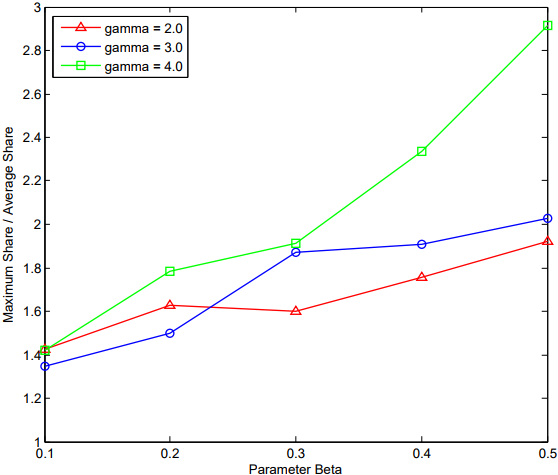} \label{cluster4maxload2} }

\subfigure[The relation among $\beta$, $\gamma$ and the standard deviation of the cluster load when the network size is decreasing quickly.] { \includegraphics[width=5cm,height=3.8cm]{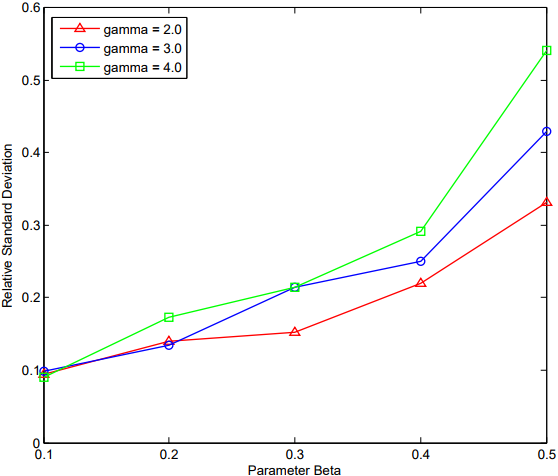} \label{cluster4standaredeviation2} }\ \ \subfigure[The relation among $\beta$, $\gamma$ and the number of metadata movement when the network size is decreasing quickly.] { \includegraphics[width=5cm,height=3.8cm]{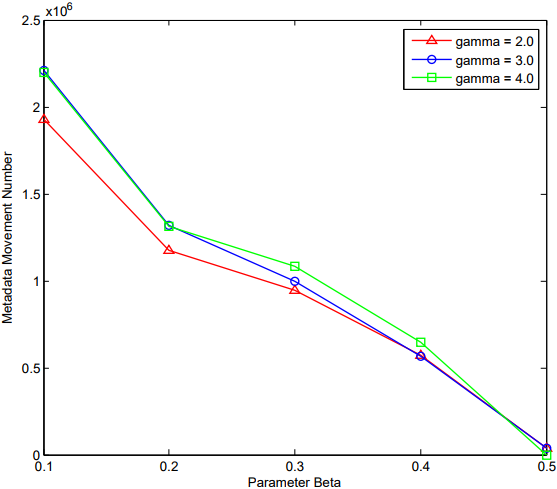} \label{cluster4move2} }

\caption{Relation among parameters  $\beta$, $\gamma$ and the effectiveness of the algorithm for the inter-cluster load balancing (I)} \end{figure*}

\begin{figure*} \centering
\subfigure[The relation among $\beta$, $\gamma$ and the ratio of the heaviest cluster  load  and average cluster load when the network size is increasing quickly.] { \includegraphics[width=4.1cm,height=3.8cm]{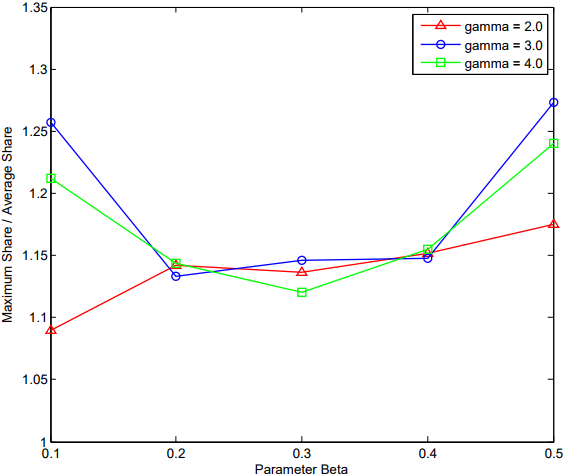} \label{cluster4maxload3} } \ \ \subfigure[The relation among $\beta$, $\gamma$ and the standard deviation of the cluster load when the network size is increasing quickly.] { \includegraphics[width=4.1cm,height=3.8cm]{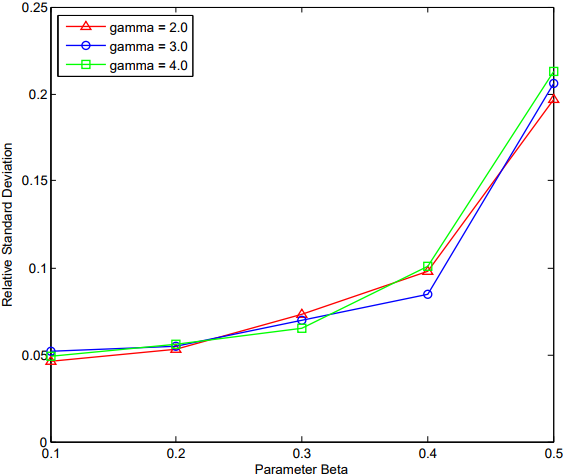} \label{cluster4standaredeviation3} }\ \ \subfigure[The relation among $\beta$, $\gamma$ and the number of metadata movement when the network size is increasing quickly.] { \includegraphics[width=4.1cm,height=3.8cm]{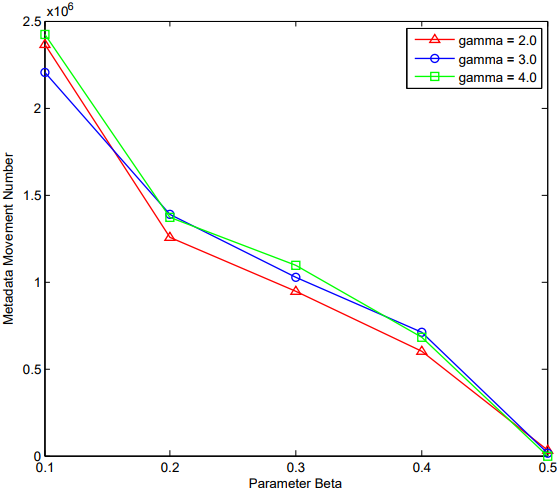} \label{cluster4move3} } \caption{Relation among parameters  $\beta$, $\gamma$ and the effectiveness of the algorithm for the inter-cluster load balancing (II)} \end{figure*}

In the inter-cluster load balancing algorithm,  $\beta$ is used to determine whether a cluster can find a neighbor cluster to transfer some loads from it, while  $\gamma$ is used to determine whether a cluster with heavy load can be split into two sub-clusters. Therefore, their different settings may have different effects on the result of the inter-cluster load balancing algorithm. In this case, we do  simulations about different settings of parameters $\beta$ and $\gamma$ in order to optimize their values. In simulations, when $\beta$ is set to  0.5, it means the cluster will not move to promote the load balance.

Fig. \ref{cluster4maxload1} - Fig. \ref{cluster4move2} show that the degrees of the cluster load balancing and metadata movement are mainly determined by  $\beta$, while  $\gamma$ will take effect, when the value of $\beta$ is so large that the cluster cannot find a neighbor cluster to transfer its load.
According to the results in Fig. \ref{cluster4maxload1} - Fig. \ref{cluster4move1}, when the proportion of nodes arrival to departure is set 1\%:1\%, meaning that MHP2P size is not subject to quick changes, $\beta$ can be valued within the interval [0.2,0.3], while $\gamma$ can be valued as 2.0. Compared with the value of $\beta$ in [0.2,0.3], that in [0.1,0.2] improves the small amplitude of the cluster load balancing by costing a large number of  extra metadata movement, and the value of $\beta$ in [0.3,0.5] sharply decreases the degree of the cluster load balancing. The value of $\gamma$ is set to 2.0 which makes the cluster with heavy load split timely without bringing any extra metadata movements on the whole.
The results in Fig. \ref{cluster4maxload2} - Fig. \ref{cluster4move2} indicates $\beta$ can be valued within the interval [0.2,0.4], while $\gamma$ can be valued as 2.0 when MHP2P size is decreasing quickly.

\begin{figure*} \centering \subfigure[The ratio of the highest load rate to average load rate under different network sizes.] { \includegraphics[width=5cm,height=4.4cm]{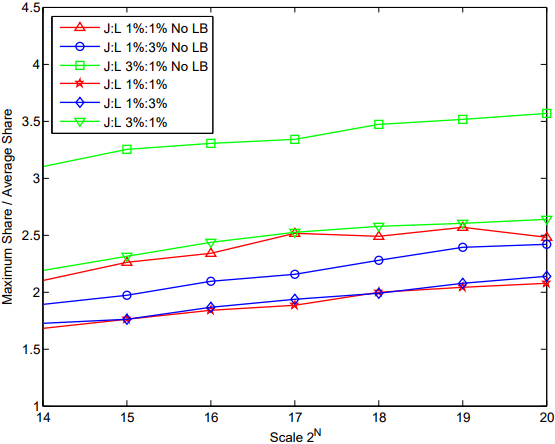} \label{Node1maxload} } \ \ \subfigure[The standard deviation of load rates under different network sizes.] { \includegraphics[width=5cm,height=4.4cm]{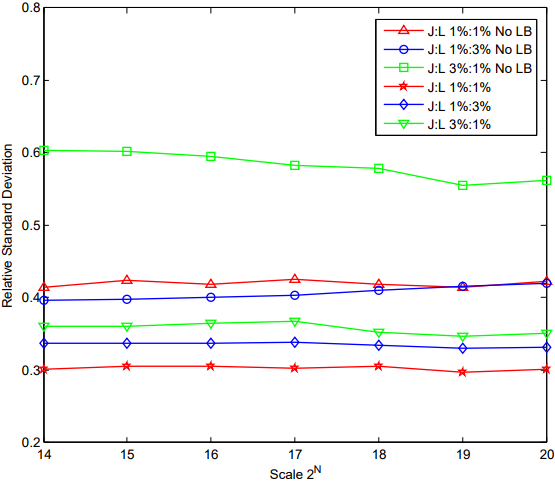} \label{Node1standaredeviation} }

\subfigure[The ratio of the highest load rate and average load rate under different numbers of metadata items.] { \includegraphics[width=5cm,height=4.4cm]{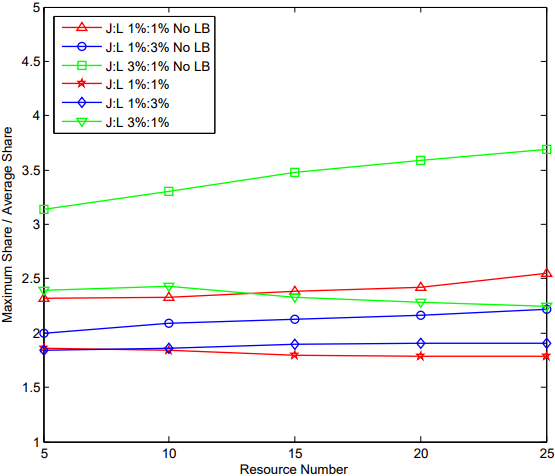} \label{Node2maxload} } \ \ \subfigure[The standard deviation of load rates under different numbers of metadata items.] { \includegraphics[width=5cm,height=4.4cm]{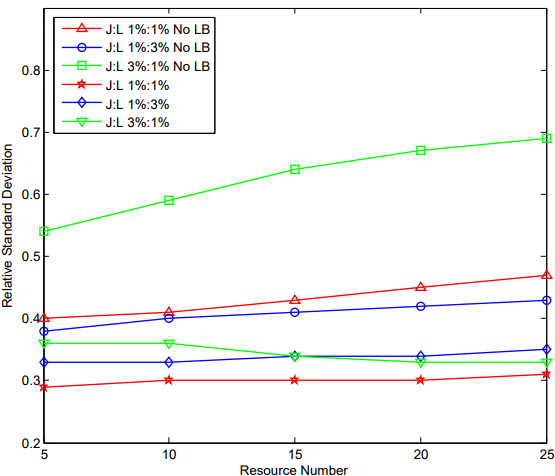} \label{Node2standaredeviation} } \caption{Simulations for the intra-cluster load balancing under different network sizes and under different numbers of metadata} \label{cluster1standaredeviationandmaxload} \end{figure*}

In Fig. \ref{cluster4maxload3} - Fig. \ref{cluster4move3}, when the network size is growing quickly, the ratio of the load of the heaviest load cluster to average cluster load and the standard deviation of the cluster load do not have obvious relationship with the value of $\beta$ in [0.1,0.4]. In this case, the process of nodes joining heavier clusters can promote the better load balancing of the clusters. Hence, $\beta$ can be set in the interval [0.4,0.5), and $\gamma$ can be set as 2.0.

According to the above simulation results, we can conclude that the inter-cluster balancing algorithm can not only decrease the cluster load with heavy load drastically, but also improve the degree of the cluster load balancing in the whole MHP2P upper layer network.

\begin{figure*} \centering \subfigure[The relation between $\alpha$ and the ratio of the highest load rate to the average load rate.] { \includegraphics[width=4.1cm,height=4.2cm]{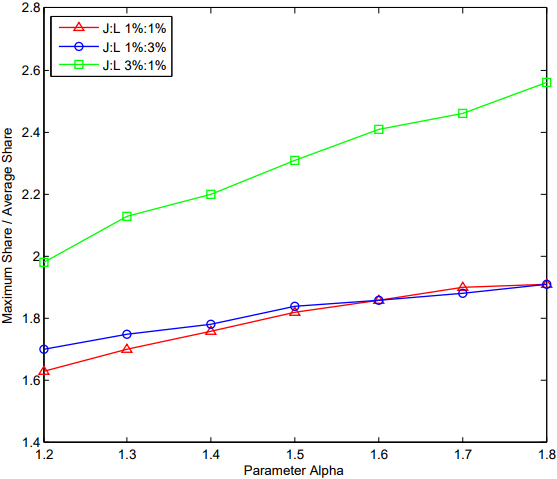} \label{Node3maxload} } \ \ \subfigure[The relation between $\alpha$ and the standard deviation of the load rate.] { \includegraphics[width=4.1cm,height=4.2cm]{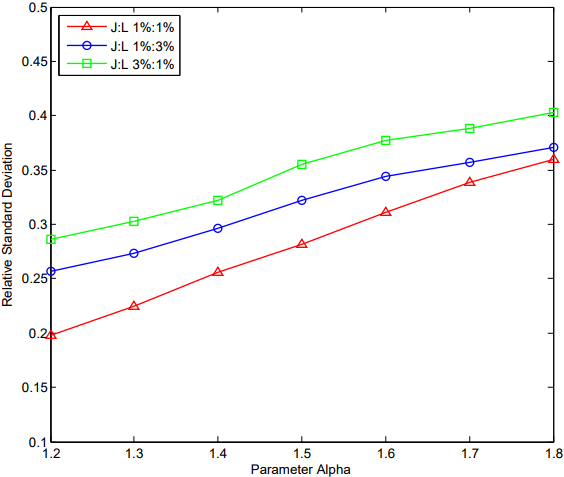} \label{Node3standaredeviation} } \ \ \subfigure[The relation between $\alpha$ and the number of metadata movements.] { \includegraphics[width=4.1cm,height=4.2cm]{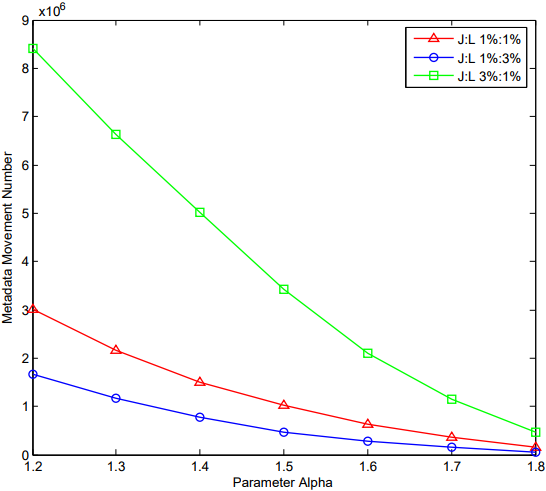} \label{Node3movement} } \caption{Relation between parameter $\alpha$ and the effectiveness of the algorithm for the intra-cluster load balancing} \label{intraclusterparameter} \end{figure*}

\subsection{Simulation of intra-cluster load balancing algorithm}

In the simulation of the intra-cluster load balancing algorithm, we mainly calculate the ratio of the highest load rate to average load rate as well as the relative standard deviation of the load rate.  The defined ratio is used to analyze whether the algorithm can relieve the node with the high load rate while the  relative standard deviation of the load rate is used to test whether the algorithm can improve the degree of the load balancing in a cluster. Since there are many clusters in the MHP2P network, the results of the ratio and relative standard deviation are the average values of all clusters.

\subsubsection{Simulations under different network sizes}

In this simulation, different network sizes from $2^{14}$ to $2^{20}$ are used to determine whether the algorithm is effective.

In Fig. \ref{Node1maxload}, an MHP2P network with the intra-cluster load balancing algorithm can keep the load rate of a node with the highest load rate about twice of the average load rate, while the value of MHP2P network without the intra-cluster load balancing algorithm can be more than 3 times in the worst case scenario. Fig. \ref{Node1standaredeviation} shows that the MHP2P network with the  algorithm  can also improve the degree of load balancing among the nodes in the same cluster. The largest network size in our simulation is in millions in order to indicate the scalability of the intra-cluster load balancing algorithm.

\subsubsection{Simulations under different numbers of metadata items}

In a cluster, different numbers of metadata items cause different load imbalance problems. Hence, we simulate the intra-cluster load balancing algorithm under different numbers of metadata items to indicate if the algorithm has any effect on the MHP2P network.  In this simulation, the average number of metadata items varies from 5 to 25.

Figs. \ref{Node2maxload} and \ref{Node2standaredeviation} show that the intra-cluster load balancing algorithm can work well under different numbers of metadata items. The algorithm can keep the ratio of the highest load rate to  average load rate below about 2.5, and the relative standard deviation is much lower than the result without it.

\subsubsection{Simulations under different settings of parameter $\alpha$}

Since the value of $\alpha$ determines the proportion of the nodes with high load rate and the proportion of those with low load rate, $\alpha$  has a big effect on the result of the intra-cluster load balancing algorithm. If  it is too high, the degree of the load balancing in a cluster is low; while if it is too low, the metadata movement is too much. Hence, we do this simulation to optimize the setting of $\alpha$. In the simulation,  $\alpha$ varies from 1.2 to 1.8.

Figs. \ref{Node3maxload} and \ref{Node3standaredeviation} show that the ratio of nodes with the highest load rate to the average load rate and standard deviation of load rate increase as  $\alpha$ increases. In Fig. \ref{Node3movement},  metadata movements decrease as $\alpha$ increases and this deceleration slows. Further, when the network size is increasing,  metadata movement is much more than that  when the network size is not increasing quickly. This is because the growth of the network size leads to the growth of metadata, and a node also needs to assign some metadata when it joins a cluster.

Fig. \ref{Node3standaredeviation} shows that when $\alpha$ is 1.2, the relative standard deviation is still greater than 0.2. Hence, if  $\alpha$ is set to 1.2, it may cause the problem of too many nodes with high load rate. In addition,   this also increases the burden on the supernode of a cluster.  $\alpha$ should hence be set in interval [1.4,1.7], and can be selected according to the requirement for the degree of the intra-cluster balance.

The result of the simulation shows that the intra-cluster load balancing algorithm  can also make the node with high load rate decrease its load rate by transferring some metadata to some nodes with low load rate. It improves the degree of the intra-cluster load balancing greatly.

\section{Conclusion}\label{sec:conc}

This paper presents a three-layer mobile hybrid hierarchical P2P model called MHP2P as a cloudlet in MEC systems.
An MEC server in the cloudlet can be any device willing to offer service.
MHP2P uses a Chord ring as the upper layer, clusters as the middle one and mobile devices as the lower one.
DHT and flooding methods employed in our model make MHP2P have high stability, scalability and efficiency.
More importantly, inter-cluster and intra-cluster load balancing schemes are provided to solve the problem of load imbalance in MHP2P. A large number of experimental simulations are conducted and the results show that the proposed schemes can significantly improve the degree of load balancing even when the model size is in millions. In the future, we will further investigate more mobile end-user based services such as gaming and VR as well as security problems with MHP2P model.

\bibliographystyle{elsarticle-num}
\bibliography{sigproc}

\end{document}